# Particle-hole symmetry breaking in the pseudogap state of Pb$_{0.55}$Bi$_{1.5}$Sr$_{1.6}$La$_{0.4}$CuO$_{6+\delta}$: A quantum-chemical perspective

Itai Panas

Energy and Materials

Department of Chemistry and Biotechnology

Chalmers University of Technology

**Abstract**

Two Bi2201 model systems are employed to demonstrate how, beside the Cu-O σ-band, a second band of purely O2p$_\pi$ character can be made to cross the Fermi level owing to its sensitivity to the local crystal field. This result is employed to explain the particle-hole symmetry breaking across the pseudo-gap recently reported by Shen and co-workers, see M. Hashimoto et al., *Nature Physics* **6**, (2010) 414. Support for a two-bands-on-a-checkerboard candidate mechanism for High-Tc superconductivity is claimed.

**Introduction**

Generic electronic structure features of the holes doped cuprate superconductors are presently being exposed at a renewed and even accelerating pace. The emerging consensus has coexistence of two energy gaps in an electronically highly inhomogeneous state of matter. The gaps are associated with the superconducting state and a pseudo-gapped state [1], respectively. The pseudo-gap function is discriminated by the fact that it displays only minor changes above and below the critical temperature for superconductivity. On-going progress is made possible by the combination of insights gained from the *real-space* Scanning Tunneling Microscopy/Spectroscopy STS/STM technique [2-8], and *reciprocal-space* Angle Resolved Photoemission Spectroscopy ARPES [9-20]. These complementary techniques have acquired impressive precision over the years, partly driven by the unsolved riddle of High-$T_C$ Superconductivity HTS found in said cuprates. Their success is proposed below to reflect fundamental complementarities inherent in the underlying physics of the HTS phenomenon.

One contemporary phenomenological perspective on the HTS phenomenology has the pseudo-gap state and the superconducting state to reflect two competing phases, see e.g. [9,10]. A complementary possible understanding is that the pseudo-gapped state signifies a segregated "pre-formed" pairs phase [11-13]. A popular conceptual microscopic framework is the Anderson Resonating Valence Bond RVB model [21-23] based on the Gutzwiller method [24] akin to the treatment by Rice and Ueda [25] of the periodic Anderson model [26] for heavy fermions. Here, a charge carrier segregated reference system, as manifested in the checkerboard structure, could be understood to act in-plane charge buffer for the hole doped AFM subsystem, allowing the SC to fine-tune the AFM doping via the Anderson hybridization term.

A complementary microscopic view point is explored by us. It is a straightforward conceptual extension to said interpretation of the Anderson RVB model but rather than solely providing hole buffering sinks, the holes segregated plaquettes become entangled by inter-plaquette interactions reflecting the sharing of pair states, i.e. Cooper pairs, as mediated by virtual magnons in the doped AFM embedding [27-29]. The rigidity of inter-plaquette entanglement is reflected in the SC gap function, while the rigidity of the virtual pair-susceptible AFM medium is reflected in the pseudo-gap function, which becomes renormalized due to the appearance of the former. This conceptual understanding is reminiscent of a spinon-holon terminology (see [21-23] again), where the spin-charge separations quality is replaced by the assumption that *neither* local spin and *nor* local space symmetry are valid local descriptors in either local AFM medium or plaquette separately. It is the fact that the AFM+Plaquette compound system *does* preserve local spin and symmetry *in conjunction with* the spatial extension of local AFM, which enforces the entangled plaquettes' ground state.

In light of the above, any new information regarding the properties of the pseudo-gap is met with outmost interest. Recently the temperature evolution of the pseudo-gap was mapped out in detail by means of ARPES in terms of signatures of the band structure at the Fermi level in the vicinity of the superconducting gap function anti-nodal direction [20]. That study succeeds in monitoring the opening of the pseudo-gap, and even more interestingly, claims to demonstrate fundamental Particle-Hole Asymmetry PHA. Phenomenological modeling based on the sufficiency of charge carrier segregation in the form of periodic checkerboard superstructures to achieve the PHA, was attempted in [20].

What adds to the drama is that the opening of a gap in a single band scenario, i.e. single-band Peierls instability [30], does not *per se* render PHA. At least a two-bands physics is required. Because conventional theory implies particle-hole symmetry conservation across the SC gap, the observed PHA across the pseudo-gap is effective in discriminating the pseudo-gapped state from the superconducting state. However, this argument is equally valid in excluding the above Peierls type charge density wave as a candidate for the pseudo-gapped state. What is required for the particle-hole asymmetry is indeed the existence of non-equivalent particle sink and source states. Such a situation may indeed be inferred from the two central dogmas in contemporary HTS theory. One is the validity of LDA band structure in providing a general basis for interpretation of the HTS phenomenology in term of the in-plane Cu-O $\sigma$-bands, see e.g. [31-33]. In these models, the required two bands for particle-hole asymmetry would comprise the upper and lower Hubbard bands, the former acting as electron sink and the latter as electron source. Assumed instability towards holes segregation would result in the checkerboard CDW observed directly by STM and indirectly as anomalous spectral broadenings in ARPES.

The purpose of the present study is to propose the complementary view to said upper/lower Hubbard bands phenomenology. Hence, sufficient conditions for the hole-doped lower Hubbard band to act as electron sink and a pure in-plane oxygen band of $O2p_\pi$ to take the role of electron source is demonstrated. Such a redistribution of holes can occur owing to innate inhomogeneous crystal fields as well as by the attenuation of the crystal field as caused by the displacement of $Sr^{2+}$ ions relative to the $CuO_2$ planes. The two model compounds $Bi_{1.5}Pb_{0.5}Sr_{1.5}La_{0.5}CuO_6$, and $Bi_2Sr_2CuO_6$, are employed here to demonstrate the two cases, respectively. In what follows, the contextual implications of our complementary understanding will first be reiterated. Second, the influence of the relative positions of the dopants $Pb^{2+}$ and $La^{3+}$ replacing $Bi^{3+}$ and $Sr^{2+}$, respectively, on the resulting band structure and partial density of states will be demonstrated. Thirdly, the corresponding effects due to displacement of $Sr^{2+}$ positions in pure $Bi_2Sr_2CuO_6$ will be demonstrated. Finally, the proposed two-bands scenario in a checkerboard framework will be employed to interpret the central observations of Shen and co-workers [20].

**A multi-band scenario for High Critical Temperature Superconductivity**

The scenario for HTS formulated by us [27-29] includes three steps. (i) At elevated temperatures mobile holes reside in the dispersive bands produced by the Cu-O σ-states. (ii) Upon cooling the charge carriers become trapped, e.g. as Zhang-Rice singlets [34], or transferred into bands of $O2p_\pi$ character. Indeed, recovery of local anti-ferromagnetic coupling [35] requires such a transfer. The opening of the pseudo-gap has two contributions, one is the development of AF coupling among $Cu3d^9$ sites, and the second is the complementary clustering of holes in "super-atom" states spanned by linear combinations of $O2p_\pi$ states. (iii) HTS emerges from a two-gaped "normal" state, such that resonant coupling of virtual holes clusters excitations and complementary virtual magnons contribute to the correlated ground state. Aspects of this understanding have been articulated in terms of a real-space analog [27] to the Bardeen Cooper Schrieffer theory, and in an equivalent two-component RVB Bose-Einstein Condensation BEC formulation [29]. The latter implies that BEC among virtual holes cluster excitations is mediated by BEC of virtual magnons. Because the one provides the coupling required for the other to condense, the corresponding two signatures (superconductivity and spin-flip resonance [36]) appear at the same temperature, i.e. at $T_C$. Our physical understanding [27-29] is similar to that of [37]. Yet, the realization of said physics is different both with regard to the detailed mechanism and in the fundamental two-bands origin of the electronic structure, subject to segregation of charge carriers. The understanding developed for the cuprates [29] was employed to formulate the superconductivity in FeSe [30], i.e. in terms of an analogous multi-band scenario to that developed for the cuprates. While such complexity is generally accepted in case of the Fe-chalcogenides and Fe-pnictides, single-band scenarios still dominate in case of the cuprates.

The report by Shen and co-workers [20] may provide the first solid ARPES based evidence in favor of a multi-band mechanism promoted by segregation in case of the HTS cuprates.

**$Bi_{1.5}Pb_{0.5}Sr_{1.5}La_{0.5}CuO_6$ band structure deconvolution**

Causes for the observations reported for $Bi_{1.5}Pb_{0.55}Sr_{1.6}La_{0.4}CuO_{6+\delta}$ in [20] are sought in the $Bi_{1.5}Pb_{0.5}Sr_{1.5}La_{0.5}CuO_6$ model system by means of spin polarized GGA PBE band structure calculations. Taking the $Bi_2Sr_2CuO_6$ crystal structure as point of departure, in what follows the influences of different structural replacements of $Sr^{2+}$ - $Bi^{3+}$ pairs by $La^{3+}$ - $Pb^{2+}$ pairs on the resulting electronic structures are demonstrated. In Figure 1, we note that there are three distinctly different positions for the 25% replacement of $Sr^{2+}$ by $La^{3+}$, and 25% replacement of $Bi^{3+}$ by $Pb^{2+}$. In all three cases, the ground state is a singlet. However, in all three cases a triplet state is only ~0.2 eV above the corresponding singlet. Figures 2 A and B depict the spin densities of the triplet states corresponding to the structures Fig. 1A and Fig. 1C, respectively. It is noted how the spin densities in the planes adjacent to $Sr^{2+}$-$La^{3+}$ ion pairs reflect a stronger crystal field than the planes exposed to $Sr^{2+}$-$Sr^{2+}$ ion pairs. Note in particular how the spin density on O in the $Sr^{2+}$-$Sr^{2+}$ bracketed planes acquire doughnut shape (superposition of $O2p_\sigma$ and $O2p_\pi$) where O in the $Sr^{2+}$-$La^{3+}$ bracketed planes display a dumbbell shaped spin density, i.e. $O2p_\sigma$, see Figure 2 again.

Hence, by studying properties of the triples state, qualitative insights are gained that are difficult to arrive at simply by looking at band structures and corresponding densities of states DOS:s of the singlet states, see Figure 3. We proceed by making connection between the spin densities depicted in Figure 2, and the triplet state α-spin Partial Densities of States for the $CuO_2$ planes subject to strong and weak crystal fields, respectively (see Figure 4). It is noted how for each structure the PDOS of each of the two planes is similar, while the Fermi level is displaced by the local crystal field (compare Figures 4 A&D, B&E, C&F) . Thus, the charge carrier inhomogeneities as represented in the spin densities of the triplet states in Figure 2 are clearly reflected in the PDOS:s of the triplet states shown in Figure 4.

Having said this, the spin density cannot be employed to read whether any charge carrier inhomogeneity is present in the singlet state because the DFT ground states are non-magnetic. However, having learned how said inhomogeneities are represented in the PDOS:s in case of the triplet states (compare Figures 2 and 4 again), in Figure 5 the PDOS:s are plotted for the structures in Figure 1 in their singlet states. Again, the PDOS:s in the vicinity of the Fermi level display clear similarities, and again it is observed how the Fermi level is "off-set" differently in planes experiencing the strong and weak crystal fields. This implies electron transfer from the $CuO_2$ planes experiencing the weak crystal fields into the planes experiencing the strong crystal fields.

We may now return to the band structure (Figure 3) and attempt to interpret it in terms of changing distance separating the $La^{3+}$ and $Pb^{2+}$ ions. The most dramatic effect is seen in the vicinity of $(\pi,\pi)$ where a $O2p_\pi$ band is shifted towards the Fermi level. The possible opening of a hole pocket centered at $(\pi,\pi)$ is enabled by employing the Cu-O σ-band as electron sink (*vide infra*).

**Employing $Bi_2Sr_2CuO_6$ to interpret $Bi_{1.5}Pb_{0.5}Sr_{1.5}La_{0.5}CuO_6$**

Also by comparing the band structure in Fig. 3 to that of native $Bi_2Sr_2CuO_6$ (Figure 6A) is it clearly seen how bands in the vicinity of $(\pi,\pi)$ along the $(0,\pi)$-$(\pi,\pi)$ direction become shifted towards the Fermi level in case of the $La^{3+}$-$Pb^{2+}$ doped samples. In addition, the apparent single Cu-O σ-band which crosses the Fermi level half way between $(\pi,\pi)$ and $(0,0)$ (see Fig. 6A again), is seen to split upon partial replacement of $Sr^{2+}$ - $Bi^{3+}$ pairs by $La^{3+}$ - $Pb^{2+}$ pairs, compare Fig. 3A and Fig. 6A. This reflects the inherently lower crystal field symmetry of the former compound. The fact that this a crystal field effect is demonstrated by displacing two

$Sr^{2+}$ ions 0.1 and 0.2 Å away from one bracketed $CuO_2$ plane, thus artificially creating a weak-strong field inhomogeneity between $CuO_2$ planes. The expected effect can be appreciated by comparing Fig. 6A to Figs. 6C, and 6E(G) at $(\pi,\pi)$-$(0,0)$, where the Cu-O σ-bands are seen to split.

A third marked effect is seen by comparing the PDOS:s of the inequivalent $CuO_2$ planes in $Bi_{1.5}Pb_{0.5}Sr_{1.5}La_{0.5}CuO_6$ in the singlet state (Figure 5) to the corresponding PDOS in $Bi_2Sr_2CuO_6$ (Figure 6B). Note the pronounced double-peak feature at ~0.3 and ~0.7 eV in Figure 6B, and how it is off-set upwards in the weak field environments (Figs. 5A-C), while in the strong-field environment it is down-shifted (Figs. 5D-E). While a similar effect is seen in the low-field $CuO_2$ PDOS:s caused by the 0.1 and 0.2 Å displacements of $Sr^{2+}$ ions, compare Figures 6B, 6D, 6F, little or no effect is seen in the strong field PDOS (compare Figs. 6B and 6F). One significant effect in the weak-field PDOS is a redistribution of electrons among Cu-O σ- states seen in a peak, which starts out at -0.2 eV (Fig. 6B) ends up at the Fermi level (Figs. 6D, and 6F). However, the most dramatic effect is seen in the O2p PDOS in the weak field plane, which piles up and sharpens at $E_F$. Interestingly though, little change is observed in the strong field $CuO_2$ PDOS (compare Figures 6B and 6H). This apparently implies that redistribution of electrons among the Cu-O σ and $O2p_\pi$ bands, as caused by the displacement of the large cations in the vicinity of the $CuO_2$ planes can occur also in the absence of inter-plane charge transfer.

Finally, it was mentioned for $Bi_{1.5}Pb_{0.5}Sr_{1.5}La_{0.5}CuO_6$ how a band of $O2p_\pi$ character may rise and touch the Fermi level in the $(0,\pi)$–$(\pi,\pi)$ direction, thus causing a hole pocket in the vicinity of $(\pi,\pi)$, see Fig. 3 again. It is gratifying to note how this hole pocket can be produced by the $Sr^{2+}$ ions displacements in $Bi_2Sr_2CuO_6$, compare Fig.6A, 6C, and 6E.

**Possible cause of Partice-hole asymmetry across the pseudo-gap**

Having thus made connection between the two employed model compounds, attention is given to the $(-\pi,\pi)-(0,\pi)-(\pi,\pi)$ segment in the Brillouin zone, which was considered experimentally in [20].

The present interpretation assumes the Fermi surface of the holes doped cuprates at T>T* to be well understood in terms of a holes doped Hubbard-Mott insulator based on a local Cu3d$^{9-\delta}$ electronic structure. Upon cooling, band shoulders at $\sim(\pm 0.1, \pi)$ appear at T~T* and become saturated at $\sim(\pm 0.2, \pi)$ for T<<T*. Here, this phenomenon is arrived at by first considering how upon cooling crystal field inhomogeneities cause the trapping of charge carriers. Indeed, the *a prori* symmetry broken crystal field may be insignificant at elevated temperatures due to the high mobility of the charge carriers at those temperatures but may become decisive below T*, and thus produce the observed temperature dependence of the ARPES signal. Secondly, this trapping acts destabilizing on the holes *a priori* residing in the Cu-O σ bands. Hence are holes partially transferred to the O2p$_\pi$ band. The destabilization of holes in Cu-O σ-bands has two origins of which one is the competing local AFM order and the second is the thermally modified crystal field destabilizing the O2p$_\pi$ to the extent that it becomes a hole sink as manifested in the opening of a hole pocket centered at $(\pi,\pi)$. This understanding is summarized in Figs. 7A and 7B, where the bands folding are representative of twice doubling of the unit cell in order to make partial connection to the checkerboard superstructure. The crystal field symmetry breaking inherent in the Bi$_{1.5}$Pb$_{0.5}$Sr$_{1.5}$La$_{0.5}$CuO$_6$ model system and caused in Bi$_2$Sr$_2$CuO$_6$ by manually displacing certain Sr$^{2+}$ ions, respectively, is schematically represented in Fig. 7B. Figure 7C repeats the central part of Fig.7B for the twice folded O2p$_\pi$ band. Qualitative connection to the ARPES band structure at T<T* is made by tracing the occupied and unoccupied electronic states in Figure 7C. Thus is the cause of the claimed particle-hole asymmetry [20] understood to result from two disjoint bands cross the Fermi in conjunction with a four-unit-cells modulated superstructure. Assuming further stabilization of the Cu-O σ-band and complementary destabilization of the $O2p_\pi$ band upon further cooling, it is indicated in Fig. 7C how the U-shaped Cu-O σ-band is made to cross the first V-shaped segment of the $O2p_\pi$ band. This renders the resulting band the additional complexity similar to that observed in [20] upon approaching the critical temperature for superconductivity.

The Kohn-Sham state corresponding to the $Bi_2Sr_2CuO_6$ $O2p_\pi$ band, as well as a proposed generic charge carrier segregated motif super-lattice in the *ab*-plane due to laterally inhomogeneous crystal field is displayed in Figure 8A and Figure 8B, respectively. The spectral broadening, taken in [20] to reflect electronic inhomogeneities, supports an interpretation along the line suggested in Figures 7 and 8. This is partly because (a) accumulation of charge carriers in the low-dispersive $O2p_\pi$ bands is expected to display instability towards holes clustering, and (b) ~25% random replacement of $Sr^{2+}$ by $La^{3+}$ and $Bi^{3+}$ by $Pb^{2+}$ is expected to cause random zero-dimensional charge carrier attractors due to the inhomogeneous crystal field.

Finally it is noted that this hole pocket of $O2p_\pi$ character centered at (π,π) has bearing on the electronic properties along the SC gap function nodal (0,0)-(π,π) direction. In particular the checkerboard superstructure is expected produce additional Fermi level crossings due to the $O2p_\pi$ band in the vicinity of both (π/2,π/2), and (π/4,π/4). This provides a complementary conceptual framework for interpretation of such observations [19,38].

In conclusion, band structures and densities of states have been presented which demonstrate how the stability of the $O2p_\pi$ band in the vicinity of (π,π) is affected by attenuation of local crystal fields in two Bi2201 model systems to the extent that this band crosses the Fermi level. This result has been employed to propose an alternative interpretation of the particle-hole asymmetry across the pseudo-gap reported by Shen and co-workers [20]. The possible relevance of such an observation for HTS was discussed in the context of a quantum chemical formulation of high-$T_C$ superconductivity.

**Computational details**

The band structure calculations employ the CASTEP [39] program package within the Material Studios framework [40]. The GGA PBE functional [41] was employed. Core electrons were described by ultra-soft pseudopotentials, O(6 electrons), Cu(11 electrons), Sr(10 electrons), La(11 electrons), Pb(14 electrons), Bi(5 electrons), employing a 340 eV cut-off energy. Summations over the Brillouin zone employed a $7 \times 7 \times 1$ Monkhorst-Pack grid[42].

# Figure Captions

**Figure 1.**

Crystal Structures of $Bi_{1.5}Pb_{0.5}Sr_{1.5}La_{0.5}CuO_6$, where the position of the $La^{3+}$ (light blue) with respect to the $Pb^{2+}$ (grey) differs.

**Figure 2.**

Spin densities corresponding to Figures 1A, and 1C. Note the doughnut shaped spin density on O reflecting both $O2p_\sigma$ and $O2p_\pi$ characters in the $CuO_2$ planes experiencing weak crystal field. This is in contrast to the CuO2 planes experiencing the strong crustal fields where the spin densities on the oxygens take a $O2p_\sigma$ dumbbell shape. In addition, in Figure 2B is seen that the apical oxygens take on some hole character, which is absent in Figure 2A.

**Figure 3.**

Densities of States and in-plane band structure $(0,0)-(0,\pi)-(\pi,\pi)-(0,0)$ corresponding to Fig. 1A (top), Fig. 1B (Centre), and Fig.1C (bottom).

**Figure 4.**

Triplet state. $\alpha$-spin PDOS in weak and strong ligand fields for the structures corresponding to Fig. 1A (top), Fig. 1B (Centre), and Fig.1C (bottom). Note how in the vicinity of the Fermi level the PDOS:s are similar in the A-D, B-E, and C-F pairs, while the Fermi level is off-set differently by the different local crystal fields. In case of A-D compare Figure 2A. In case of C-F, compare Figure 2B.

**Figure 5.**

Same as in Figure 4, but for the ground state singlet state. PDOS in weak and strong ligand fields for the structures corresponding to Fig. 1A (top), Fig. 1B (Centre), and Fig.1C (bottom). Note again how in the vicinity of the Fermi level the PDOS:s are similar in the A-D, B-E, and C-F pairs, while the Fermi level is off-set differently by the different local crystal fields..

**Figure 6**

(A) Band structure, and (B) $CuO_2$ PDOS for $Bi_2Sr_2CuO_6$ (compare Figure 3). Also displayed are the band structure (C) and PDOS (D) for $CuO_2$ planes in $Bi_2Sr_2CuO_6$ caused to experience artificially weak field (0.1 Å displaced $Sr^{2+}$ ions). (E) and (F) are same as (C) and (D) but with 0.2 Å $Sr^{2+}$ displacements. (G) is the same as in (E), while (H) is PDOS for the $CuO_2$ plane in the modified $Bi_2Sr_2CuO_6$ structure, which experiences an unchanged local field (see text).

**Figure 7**

(A) Section probed by ARPES [20] at T>T*. The Cu-O σ conduction band (red), and the $O2p_\pi$ band (blue). Twice folded $O2p_\pi$ band (green) assumes a four unit cells electronic modulation. (B) Symmetry breaking due to inequivalent $CuO_2$ caused by the inhomogeneous crystal field (T<T*). (C) Same as B. Stabilization of the Cu-O σ-band (dashed red) causes transfer of holes into the twice folded $O2p_\pi$ band (dashed green). Effective occupied (blue) and virtual (purple) states are indicated to make connection to [20]. (D) At T<<T* the Cu-O σ-band (dashed red) is suggested to cross the upper V-shaped segment of the twice folded $O2p_\pi$ band (dashed green) rendering the resulting band additional structural features (cf. [20] again).

**Figure 8**

(A) Kohn-Sham state at the Fermi level of $O2p_\pi$ character (see Fig.6E and 6F). (B) Generic spin density inhomogeneity in a Hg1201 super-cell due to c-axis displacement of central $Ba^{2+}$ cation. A $4a_0$ x $4a_0$ super-cell at 25% hole doping is assumed. Note the exclusive $O2p_\pi$ character (dumbbells) of the spin density on the oxygen atoms, as well as the inhomogeneity among these within the super-cell.

**Figure 1**

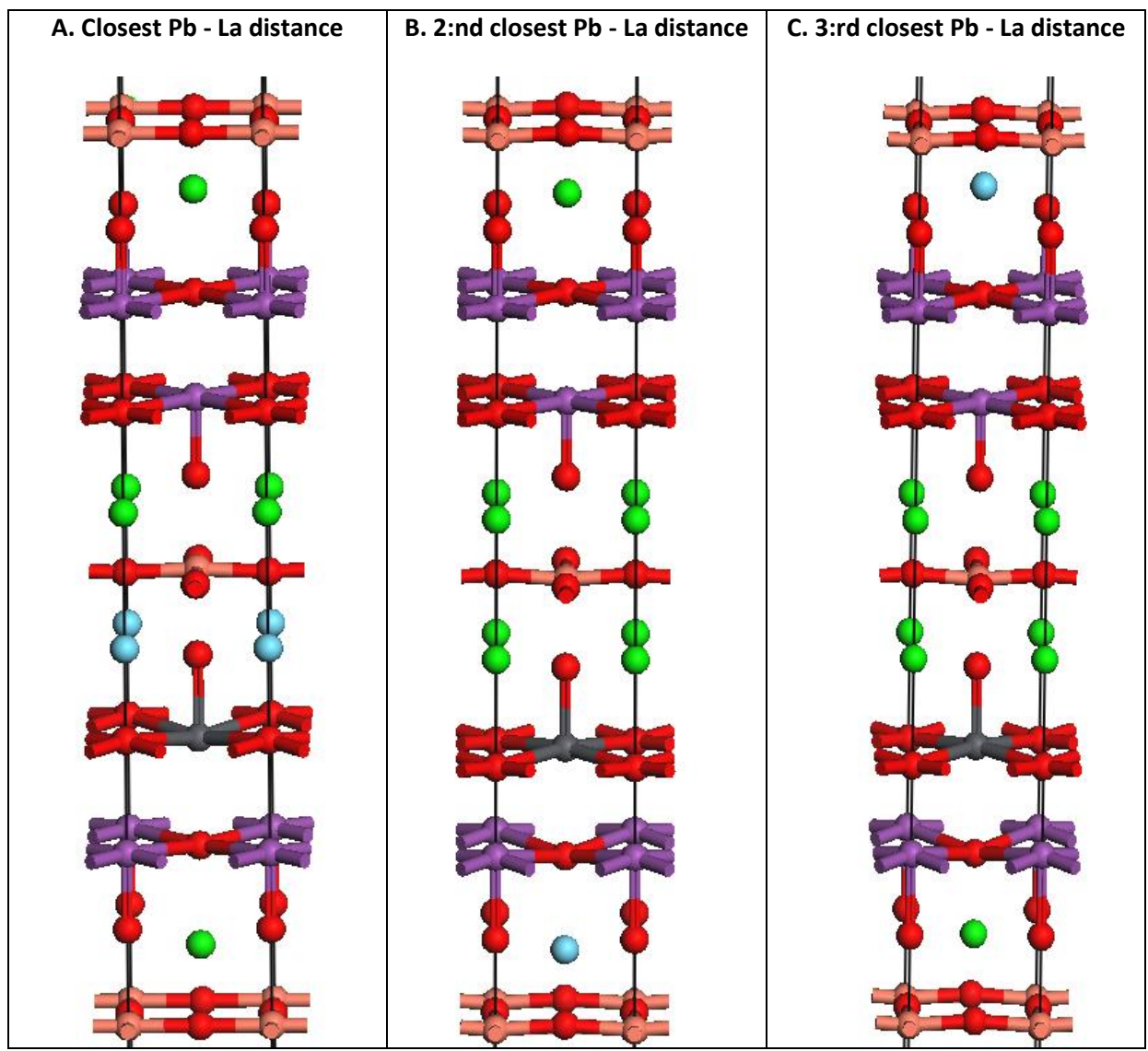

**Figure 2.A.**

# Closest Pb - La distance

**Weak field CuO$_2$ plane (top)**

**Strong field CuO$_2$ plane (bottom)**

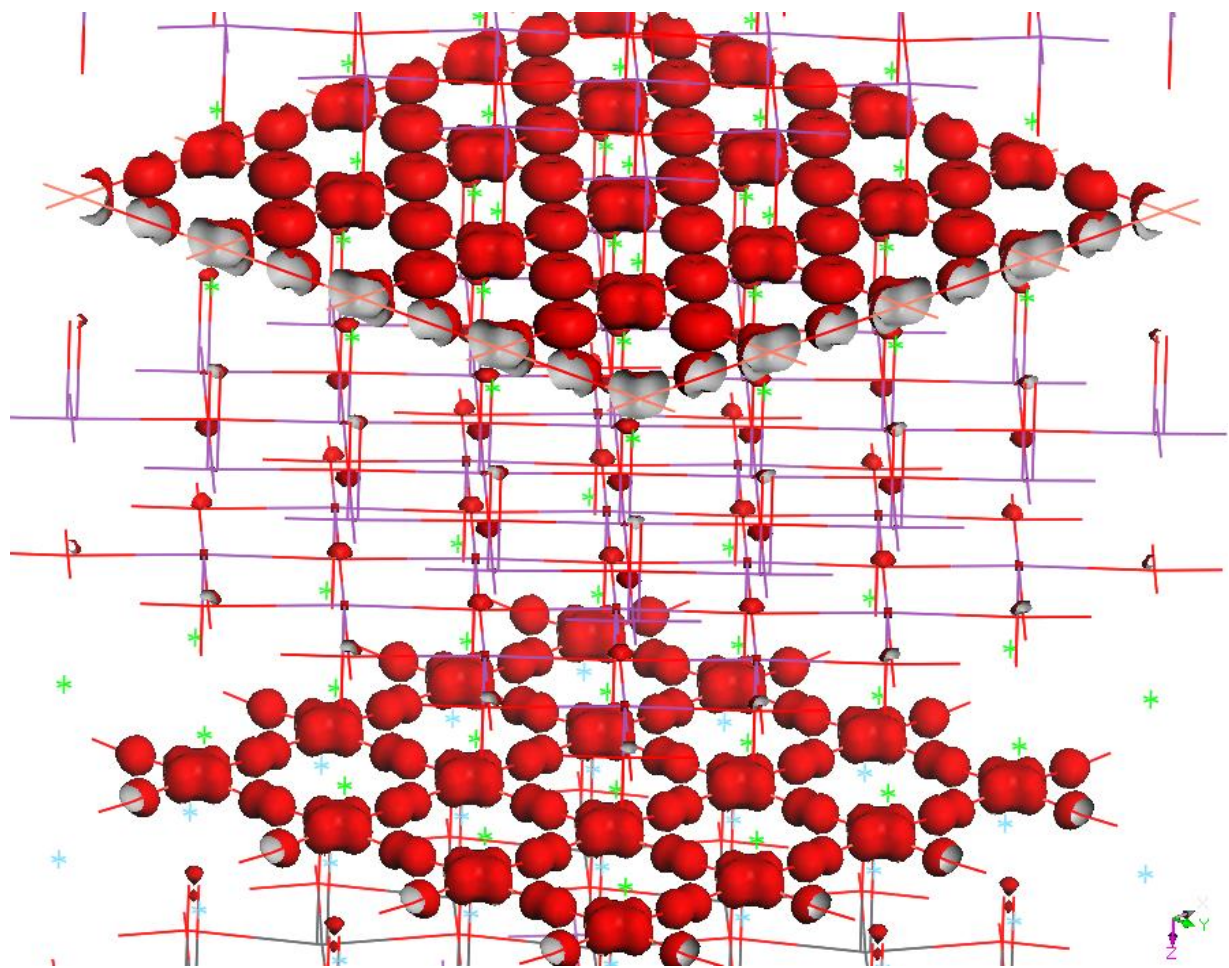

**Figure 2.B.**

### 3:rd closest Pb - La distance

**Weak field CuO$_2$ plane (top)**

**Strong field CuO$_2$ plane (bottom)**

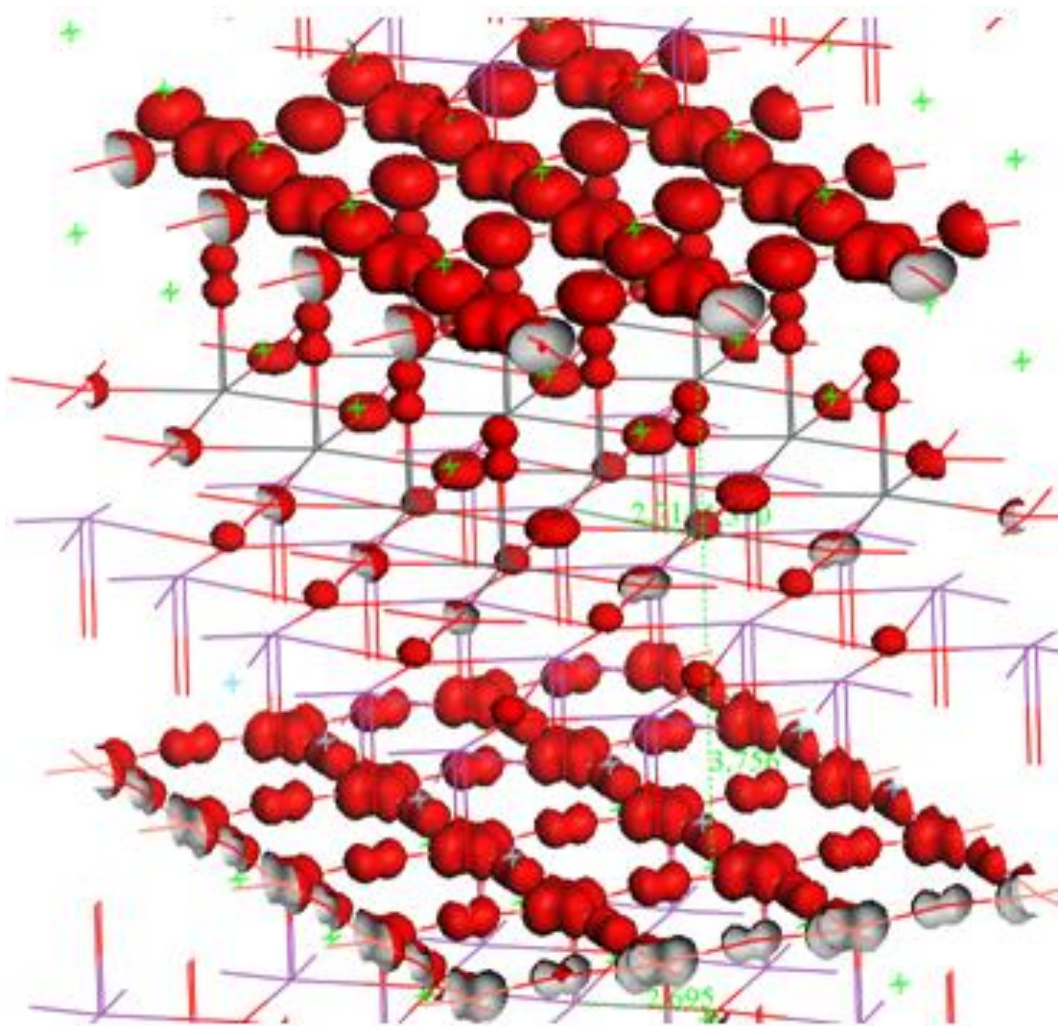

**Figure 3.**

Singlet

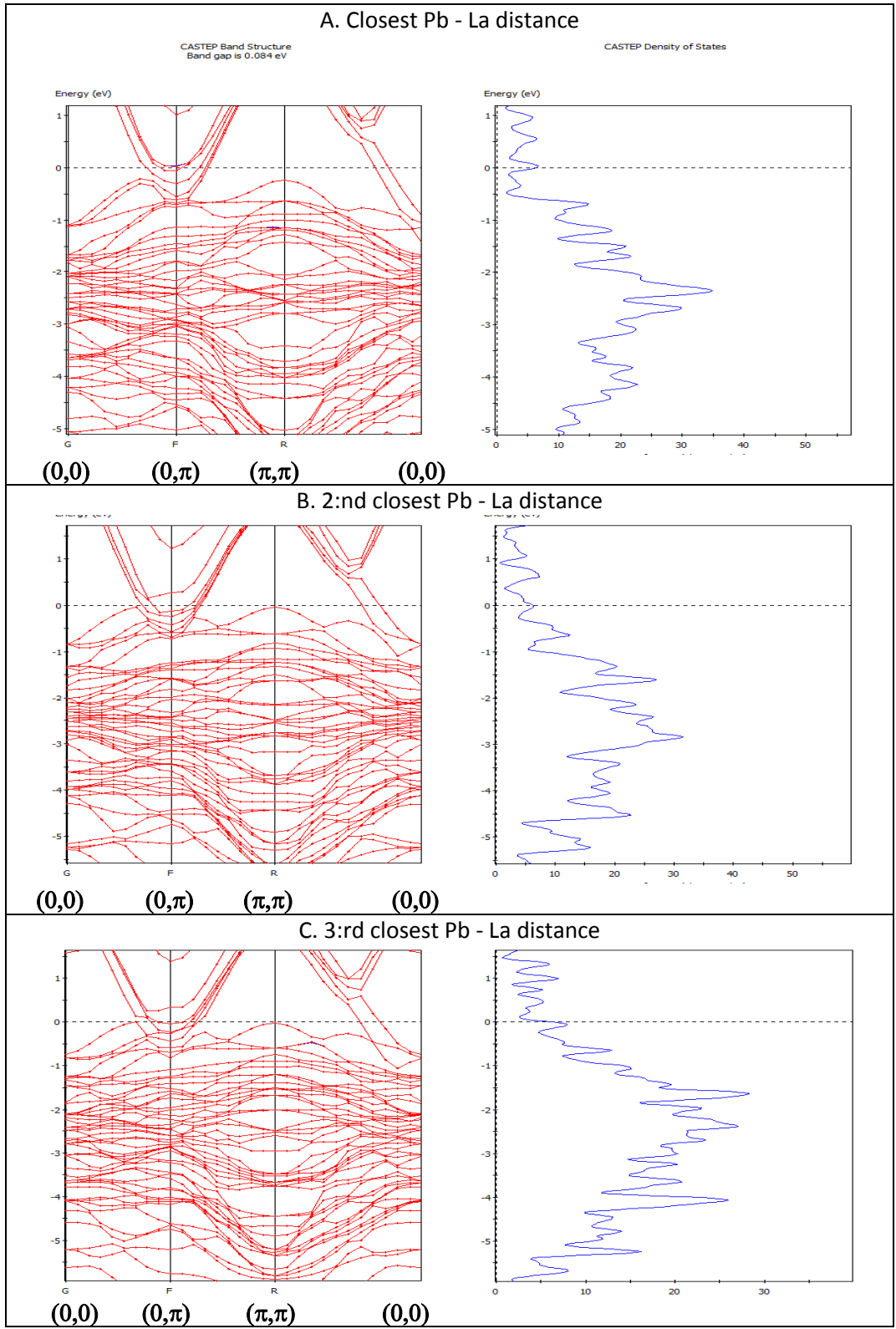

**Figure 4.**

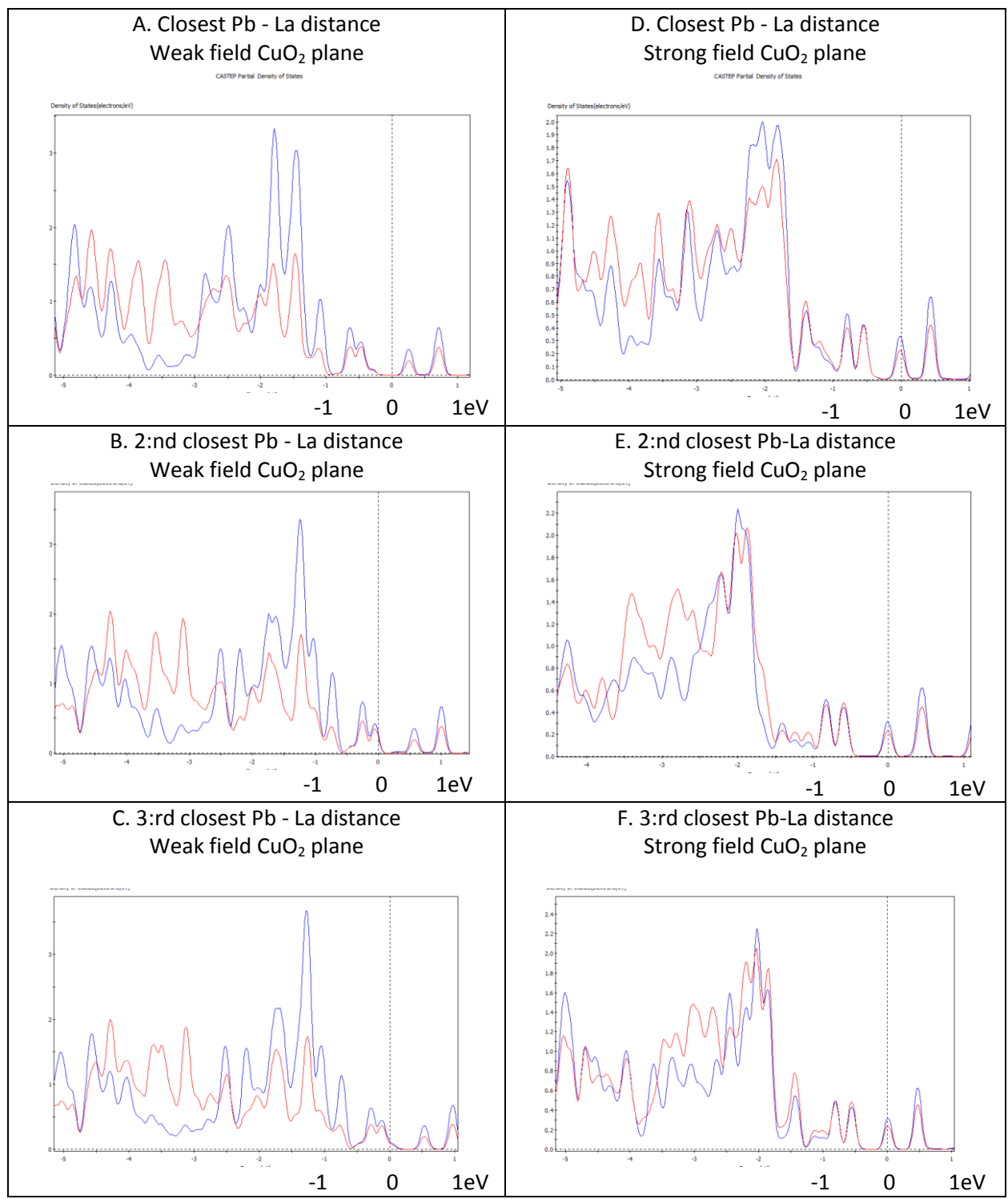

**Figure 5**

Singlet

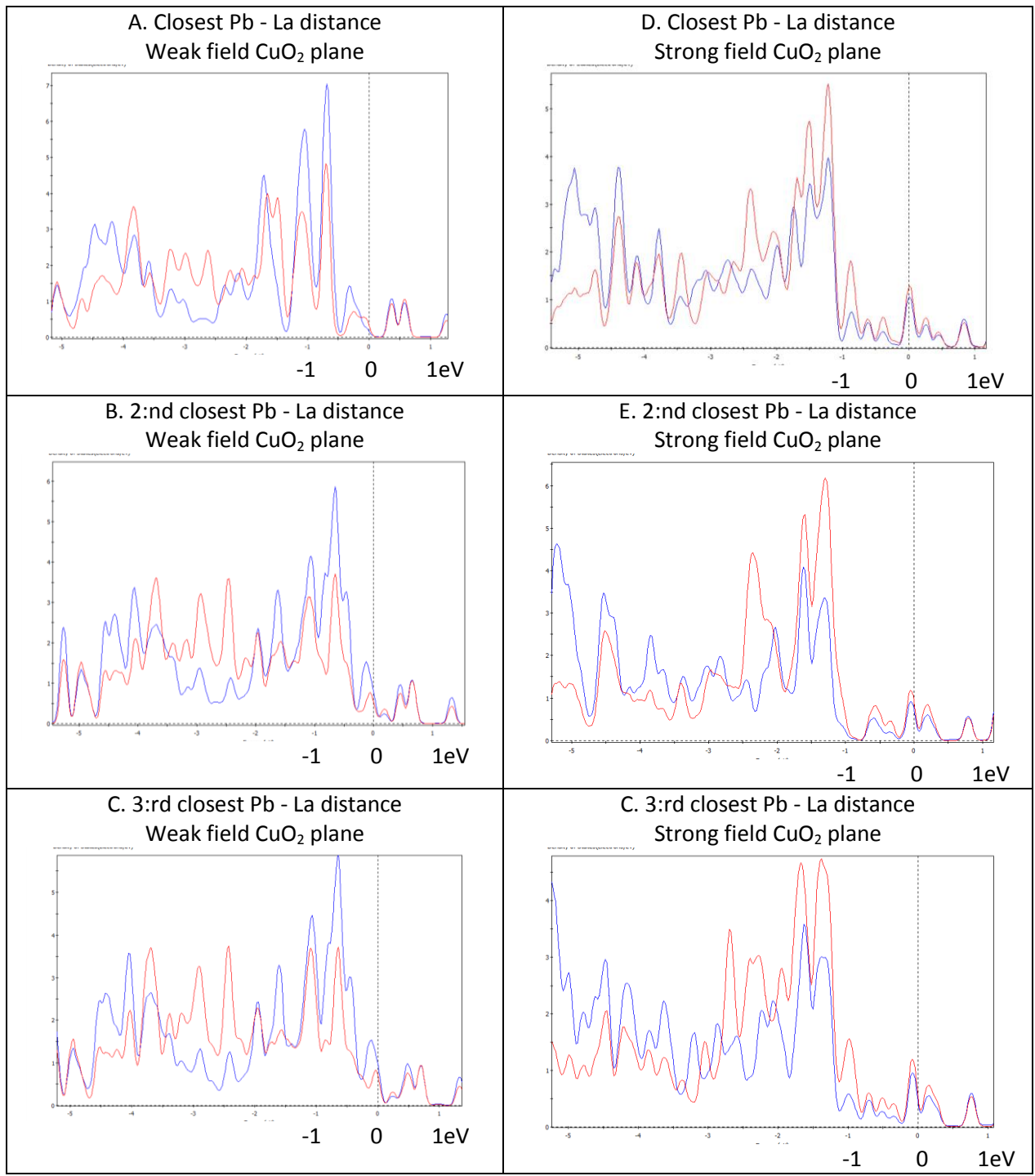

**Figure 6**

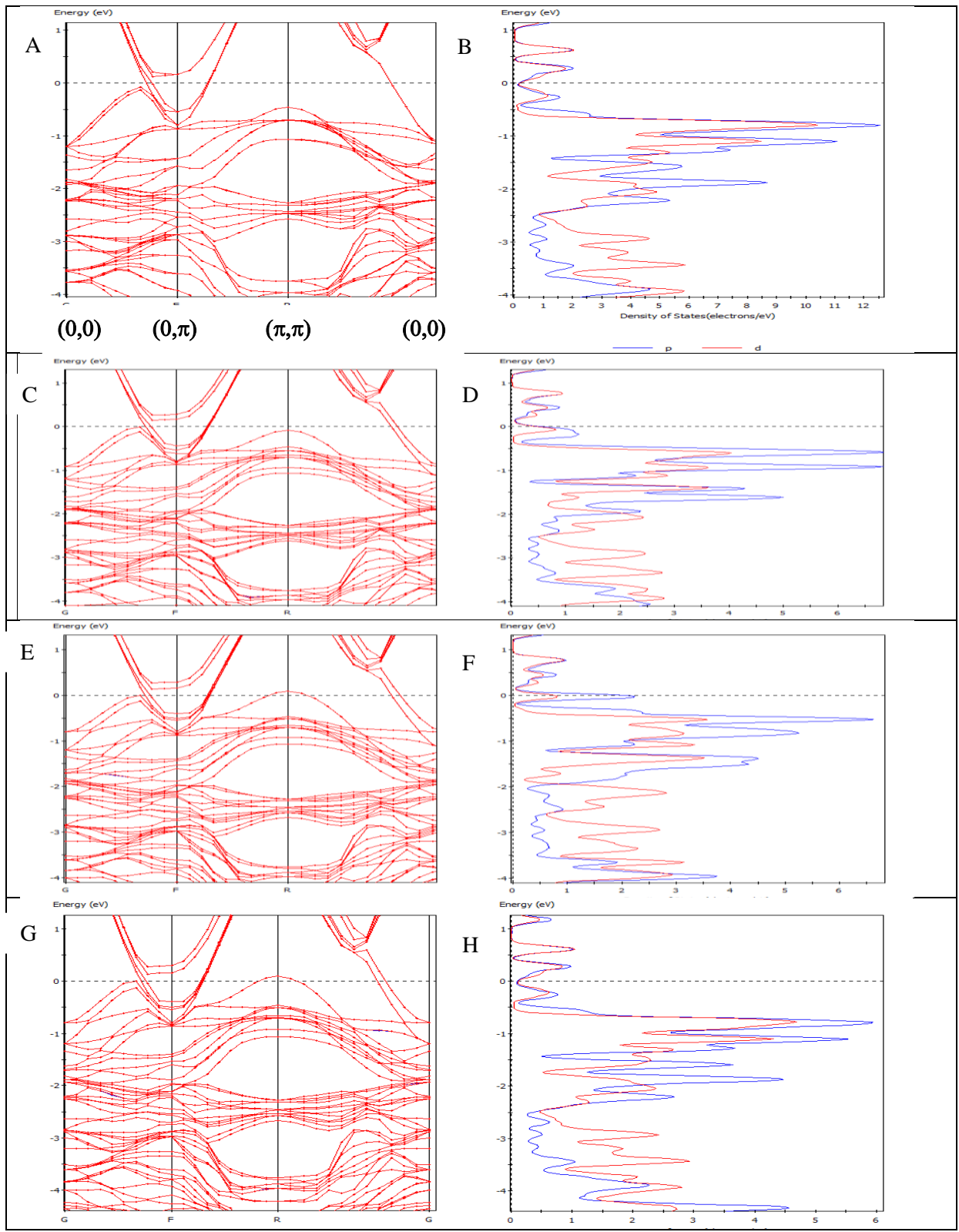

Figure 7

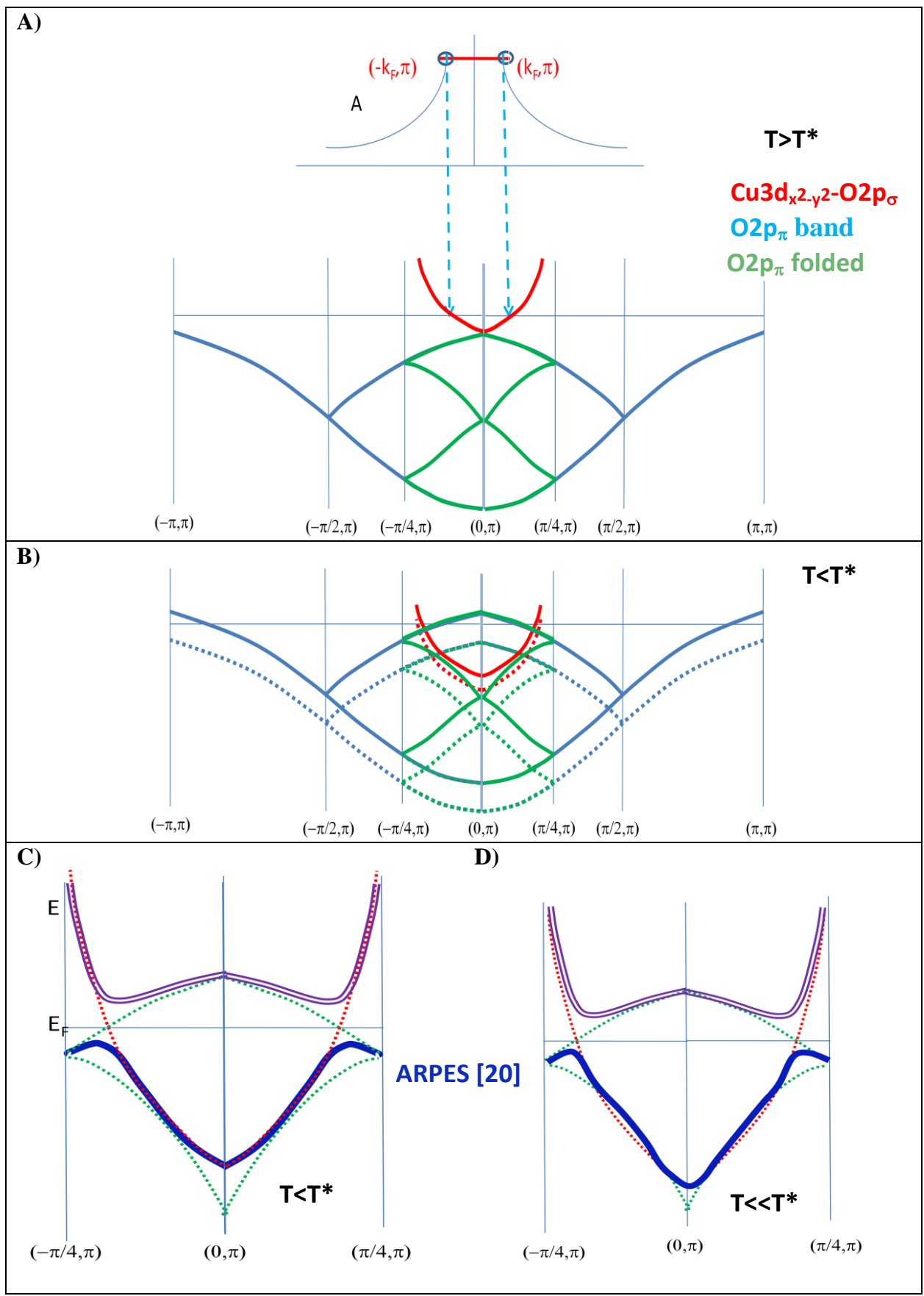

**Figure 8**

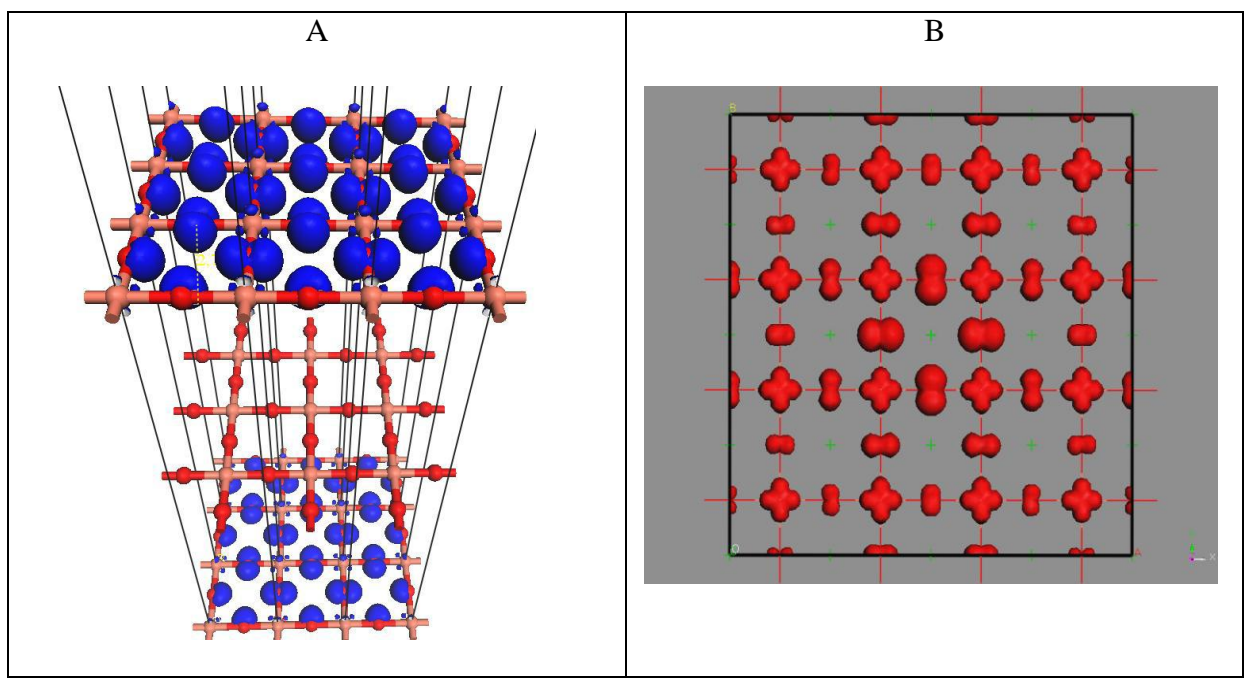